\documentclass[12pt,tightenlines,superscriptaddress,nofootinbib]{revtex4}
\usepackage{latexsym}
\usepackage{graphicx}
\usepackage{url}
\usepackage{amssymb,amsmath,amsfonts,amstext,graphics, multirow}
\usepackage{graphicx}
\usepackage{epstopdf}
\usepackage{url}
\usepackage{appendix}
\usepackage{color,slashed}
\usepackage{verbatim}

\newcommand{\gsim}{\lower.7ex\hbox{$\;\stackrel{\textstyle>}{\sim}\;$}}
\newcommand{\lsim}{\lower.7ex\hbox{$\;\stackrel{\textstyle<}{\sim}\;$}}

\def\OO{{\cal O}}

\newcommand{\TeV}{\,\mathrm{TeV}}
\newcommand{\GeV}{\,\mathrm{GeV}}

\newcommand{\met}{\ensuremath{E\!\!\!\!/_T}}
\newcommand{\mht}{\ensuremath{H\!\!\!\!/_T}}
\newcommand{\MET}{\ensuremath{\met}}

\newcommand{\bef}{\begin{figure}[htbp]\begin{center}}
\newcommand{\eef}{\end{center}\end{figure}}

\newcommand{\bea}{\begin{eqnarray}}
\newcommand{\eea}{\end{eqnarray}}

\begin{document}

\pagestyle{plain}
\title{Study of LHC Searches for a Lepton and Many Jets}

\author{Mariangela Lisanti}
\email{mlisanti@princeton.edu}
\affiliation{Princeton Center for Theoretical Science,
Princeton, NJ 08542, USA}

\author{Philip Schuster}
\email{pschuster@perimeterinstitute.ca}
\affiliation{Perimeter Institute for Theoretical Physics,
Ontario, Canada, N2L 2Y5 }
\affiliation{Institute for Advanced Study, Princeton,
New Jersey 08540, USA}

\author{Matthew Strassler}
\email{strassler@physics.rutgers.edu}
\affiliation{Rutgers University, New Brunswick, New Jersey 08854, USA}

\author{Natalia Toro}
\email{ntoro@perimeterinstitute.ca}
\affiliation{Perimeter Institute for Theoretical Physics,
Ontario, Canada, N2L 2Y5 }
\affiliation{Institute for Advanced Study, Princeton,
New Jersey 08540, USA}
\date{\today}

\begin{abstract}
Searches for new physics in high-multiplicity events with little or no missing energy are an important component of the LHC program, complementary
to analyses that rely on missing energy.  We consider the potential
reach of searches for events with a lepton and six or more jets, and
show they can provide increased sensitivity to 
many supersymmetric and exotic models that would not be detected through
standard missing-energy analyses.  Among these are supersymmetric
models with gauge mediation, R-parity violation, and light hidden
sectors. Moreover, ATLAS and CMS measurements suggest the primary background in this channel 
is from $t\bar t$, rather than $W$+jets or QCD, which reduces the complexity of background modeling necessary for such a search. 
We also comment on related searches where the lepton is
replaced with another visible object, such as a $Z$ boson.
\end{abstract}

\maketitle

\section{Introduction}

Most searches for  supersymmetry (SUSY) at hadron colliders focus on signatures with missing transverse energy ($\MET$).
These signatures are motivated by theories with stable, invisible dark matter candidates or long-lived neutral particles, such as the lightest superpartner (LSP).  In the first 36 pb$^{-1}$ of LHC data, searches for SUSY in mainly hadronic channels imposed minimum missing energy requirements ranging from 100 to 250 GeV \cite{Aad:2011xm, Collaboration:2011xk, Aad:2011ks, daCosta:2011qk, Aad:2011hh,Khachatryan:2011tk,Chatrchyan:2011bz,Chatrchyan:2011wb,Chatrchyan:2011ah}.
Demanding significant $\MET$ is a powerful strategy for rejecting Standard Model (SM) backgrounds, but sacrifices sensitivity to a variety of Beyond-the-Standard-Model (BSM) signals with smaller $\MET$, including many types of weak-scale supersymmetric models. 
Many of these theories are also unobservable in existing exotica searches, which tend to focus on few-particle resonances.
In this work, we consider a search that complements current high-$\MET$ hadronic analyses by providing sensitivity to models with  little or no missing energy.

Large classes of SUSY and exotica models with low $\MET$ share characteristic features, including
(1) high object multiplicity, particularly in scenarios where long decay chains
 deplete $\MET$, 
 (2) high $\sum |p_T|$ of reconstructed objects, associated with the produced-particle mass scale, and 
 (3) strong-interaction production cross sections, with modest suppression of lepton or electroweak gauge boson emission in decay chains. We will show that searches that take into account all three of these features have an important role to play at the LHC.  Even for models with large $\MET$, such searches can have sensitivity comparable to existing jets+$\MET$ analyses, and when effects significantly reduce the $\MET$ signal, they can retain and often gain sensitivity.  In this sense, they provide excellent complementarity to the $\MET$-based searches, covering additional large domains of parameter space.
  
\begin{figure}[bt]
\includegraphics[width=0.4\textwidth]{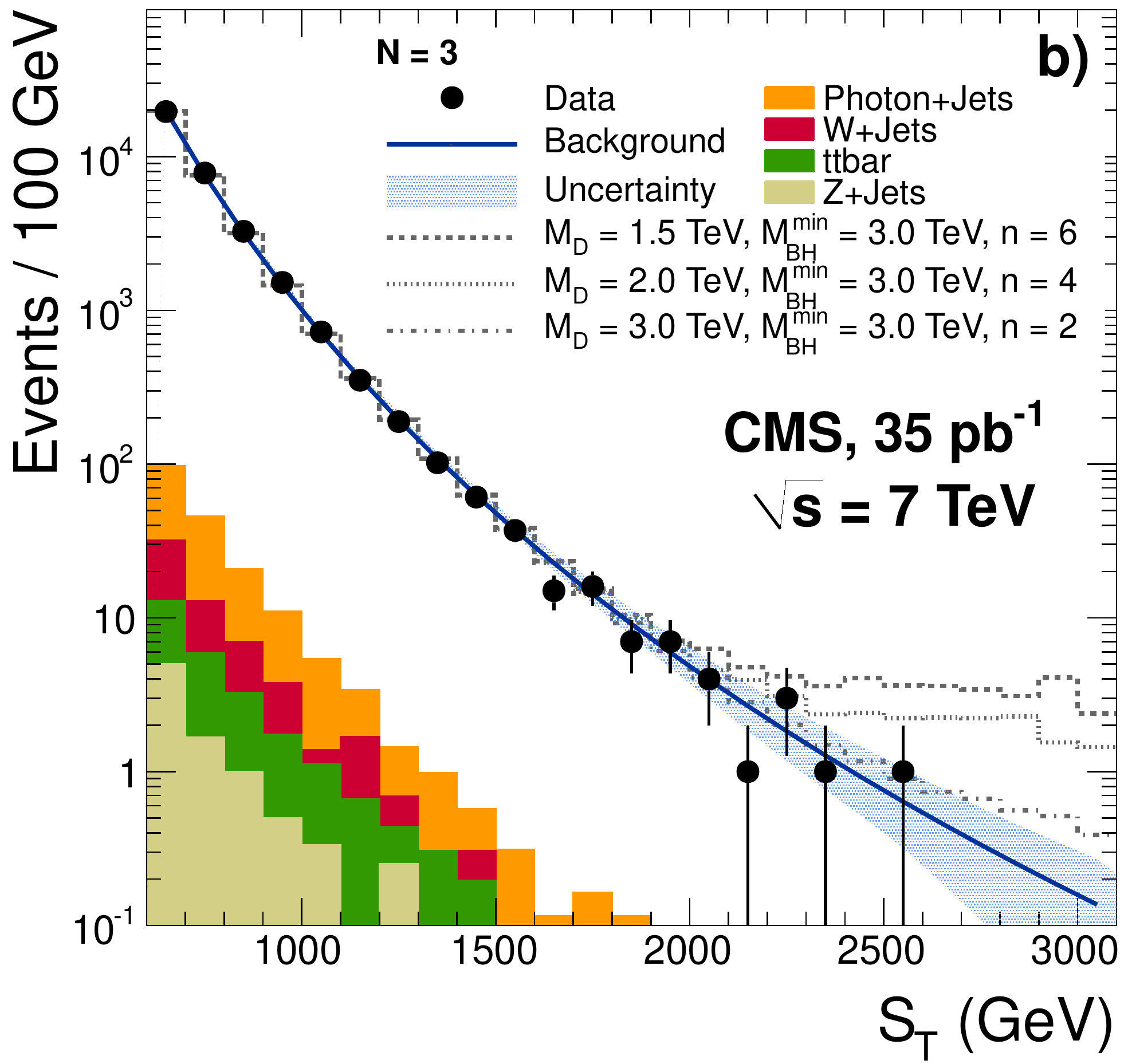}	
\caption{\label{bhPlots} 
Plot from CMS \cite{Khachatryan:2010wx} of the Standard Model $S_T$ distribution for events with precisely three reconstructed objects of $p_T > 50$.  $S_T$ is as defined in Eq.~(\ref{STdefn}).  Note the ratio of the electroweak and top pair rates (colored histograms) to the QCD-dominated total $S_T$ distribution.}
\end{figure}

The approach that we will consider involves searching for events with a visible non-jet object(s), a high multiplicity of jets, and $S_T$ above a threshold of ${\cal O}(\TeV)$, where
\begin{equation}\label{STdefn}
S_T \equiv \met + \sum_{\begin{smallmatrix} \text{visible}\\ \text{objects} \end{smallmatrix}} |p_T| \ .
\end{equation}
Such a strategy is certainly not new.  
High object-multiplicity with large $S_T$ is
already used in TeV-scale black hole searches
\cite{BHATLAS,BH3ATLAS,Khachatryan:2010wx}, where the signal is
expected to dominate at $S_T \gtrsim 2$ TeV, as seen in
Fig.~\ref{bhPlots}. In this case, QCD multi-jets is the dominant background, orders
of magnitude larger than electroweak and top production, and to remove
it requires an $S_T$ cut of multiple TeV.  While such a cut is reasonable for black
holes, whose partonic cross-section grows with energy and whose signal
peaks at ultra-high $S_T$, it is too extreme for more general low-$\MET$,
high-multiplicity new physics scenarios.  We will consider a complementary approach, 
reducing the QCD background by requiring at least one lepton or photon, 
and thus taking advantage of feature (3) above --- the relative ease of
producing electroweak bosons or leptons in new physics events.  In this way,
one gains sensitivity to signals peaked at more moderate $S_T\sim 1$ TeV.

Searches that are much closer to what we advocate include the case of a leptonic $Z$ accompanied by many
jets, which has been studied at the Tevatron in \cite{Aaltonen:2007je}.  Similar strategies have been used in broad model-independent searches, such as SLEUTH at the Tevatron~\cite{Abbott:2000gx, Abbott:2001ke, Aaltonen:2007dg, Aaltonen:2008vt, Aaltonen:2007kp}, which scanned through many different final states for excesses in the high-$S_T$ tail.  The MUSiC analysis at CMS also involves a fairly exhaustive scan seeking discrepancies between data and Monte Carlo in high $S_T$ distributions~\cite{OldMusicPaper,NewMusicPaper}.   Other black-hole searches have examined top~\cite{ATLAS-CONF-2011-070} or di-lepton final states~\cite{BH2ATLAS, BH2CMS}.  However, as yet there does not seem to exist a comprehensive program of dedicated high-multiplicity, high-$S_T$ searches in samples with leptons and/or photons.  Our aim is to advocate for such a program, by stressing the robustness and scientific importance of searches of this type.

In this article, we do this by focusing on a specific case, which we
believe to be the most powerful of the set -- searches requiring a
lepton, accompanied by six or more jets
above a $p_T$ threshold, and by a small amount of $m_T$ (as might be expected from a $W$) but with no explicit
$\MET$ cut.  This search strategy has a high efficiency
for many signals, and it appears to have a relatively simple background, dominated, for suitable cuts, by $t\bar t$ plus a small number of extra
jets. 

The following section expands on our motivation, describing how low-$\met$, 
high-multiplicity signals easily arise in SUSY and in other new-physics models.  
We discuss backgrounds for a lepton-plus-many-jet search in \S\ref{sec:backgrounds}.
In \S\ref{sec:illustrative}, we expand on the features of one family of supersymmetric models, focusing in \S\ref{ssec:35pb} on the sensitivity that might be achievable with further analysis of the 36 pb$^{-1}$ of 2010 data. 
We argue in \S\ref{ssec:higherLumi} that an expanded analysis in 2011 data will significantly enhance this sensitivity, even accounting for increases in $p_T$ thresholds.  Cross-checks that could help test whether an observed excess is due to mismodeling of the top background are suggested in \S\ref{sec:commentsOnCrossChecks}. We extend our discussion to other models in \S\ref{sec:otherModels}, thereby demonstrating that our strategy is widely applicable, and conclude in \S\ref{sec:conclusion} with a few comments.

\section{Motivation for the Multi-Jets + X Strategy}
\label{ssec:introMotivate}

It is quite possible that new physics will first manifest itself as a large signal with many 
jets, little or no missing energy, and an occasional lepton, photon, or other relatively rare visible object.  In this section we review some mechanisms by which this may naturally occur.  While this possibility may be realized in
a broad set of theories, including models with extra dimensions~\cite{Appelquist:2000nn}, Higgs and heavy-flavor  compositeness~\cite{compositeness,GregoireKatz} and a Little Higgs~\cite{ArkaniHamed:2001nc, ArkaniHamed:2002qx, Katz:2003sn}, we take supersymmetry as our main example, and focus on the case where the rare object is a lepton. We will see at the end of this section how the use of a lepton-plus-many-jet search strategy can recover sensitivity to models whose $\MET$ signal is too low to be found by standard means.

A generic signature of supersymmetry involves jets and missing energy.
By $SU(3)$ color conservation, jets always accompany strong production of squarks and
gluinos.   If R-parity is exact, and
if the supersymmetric spectrum is minimal with the gravitino heavier
than the lightest superpartner, then the lightest SM superpartner is
also the lightest R-parity-odd particle --- the ``LSP'' --- and is
stable.  If, in addition, the LSP is produced with moderately high
momentum, 
then a high-$\MET$ search is very effective.  
However, the assumptions underlying the high-$\MET$ strategy are quite
model-dependent.  Several effects can lead to
a reduction in the $\MET$ signal, and a corresponding increase in the
jet multiplicity:
\begin{itemize}
\item
{\bf Decay of the lightest SM superpartner to a partly-visible final state.}  
The most obvious feature that can
reduce the $\MET$ signal is a change in the structure of the
electroweak decay topology.  For instance, in low-scale gauge-mediated
SUSY breaking, the gravitino is the lightest R-parity-odd
particle. The lightest SM superpartner is the next-to-lightest
R-parity-odd particle (the ``NLSP''), and can decay to a gravitino in
association with its partner.  A
neutralino NLSP can decay to a photon, $Z$ or Higgs boson.  Many of the 
other possible choices for the NLSP often produce jets and/or taus.
Compared to similar models with no such decay, the
$\MET$ in the signal is reduced, and replaced with other objects,
which often include two or more jets.  
Among numerous other classes of examples, the next-to-minimal supersymmetric Standard
Model (NMSSM) often allows the
lightest standard-model superpartner to decay into a mostly singlet
Higgs-like scalar and its superpartner, with the scalar in turn
decaying to jets (often $b$'s).  In SUSY-like models,
such as universal or partly-universal extra dimensions with a
KK-parity, or Little Higgs with a T-parity, similar considerations apply.
\item
{\bf Cascade decays or squeezed spectrum.}
Even within the minimal supersymmetric Standard Model (MSSM), the kinematics of the decay topology can
be unfavorable for $\MET$-based searches. This happens when the spectrum
is modestly squeezed, or when cascade decays into $W$ or $Z$ carry off energy,
leaving the LSP with lower $p_T$ and thus reducing the $\MET$ signal.
Again, analogous issues can affect other SUSY-like models, such as universal
extra dimensions.
\item
{\bf Weakly-broken global symmetries (e.g., R-parity violation.)}
If R-parity is sufficiently violated, the $\MET$ signal in a SUSY model
can be almost entirely lost, as the LSP decays into two or three jets.
The same effect can occur in a Little Higgs model, where T-parity
plays a similar role to R-parity, and may be violated in some cases by
an anomaly \cite{Tanomaly}. The KK-symmetries of an extra-dimensional model
may also suffer some amount of violation.
\item {\bf Top-rich signals.} 
SUSY models with
light stops or sbottoms, or with light Higgsinos, are likely to produce a $t$ and a $\bar t$ in a large fraction
of the events,  if kinematics allow.  The same applies for many models of strong dynamics at the
electroweak scale, to which the top quark often couples more strongly
than other quarks.  Compared to a model that produces other quarks,
one that produces a $t$ and a $\bar t$ will have two to four extra jets (and often a lepton) in the final state.
\end{itemize}

\begin{figure}[tb]
\includegraphics[width=0.45\textwidth]{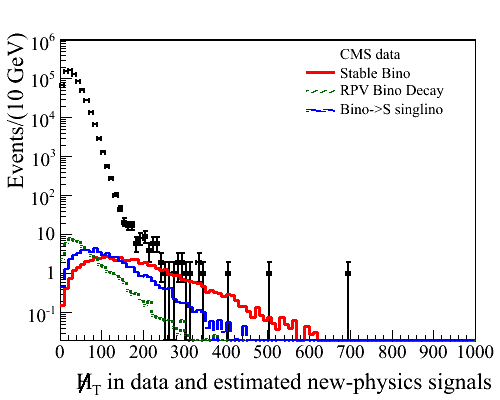}
\caption{
\label{basicMET} 
Missing energy distributions for three variations on the low-mass benchmark model described in \S \ref{sec:illustrative}, in which the bino is either a stable LSP (red solid curve), decays to an NMSSM-like singlet and singlino, followed
by singlet decay to two jets  (green dotted curve),  or decays through R-parity violation to three quarks (blue dashed).  A lepton veto, angular cuts, and $H_T>300$ GeV have been imposed (similar to the base selection of \cite{Collaboration:2011ida}).  Shown for
comparison is the data from \cite{Collaboration:2011ida}, which agree with the data-driven estimate of Standard Model background.}
\end{figure}

Meanwhile, often independently of whether the $\MET$ signal is small or large, leptons and/or photons commonly appear in the cascade decays of the colored particles.  The case of a single lepton is of particular interest, because leptons often arise as daughters of $W$ or $Z$ bosons emitted in colored-particle decays, of lepton partners (such as sleptons), or of top quarks.  This motivates us to focus our attention in this work on the potential of a lepton-plus-many-jet search.

\begin{figure}[tb]
\includegraphics[width=0.45\textwidth]{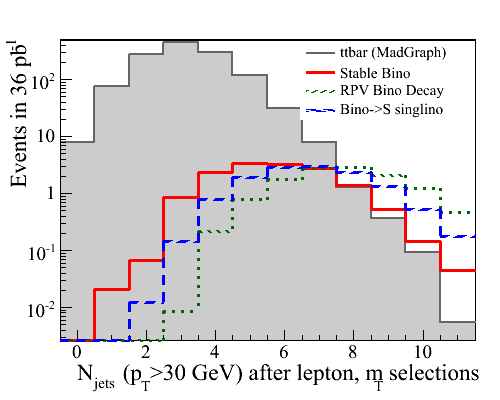}
\includegraphics[width=0.45\textwidth]{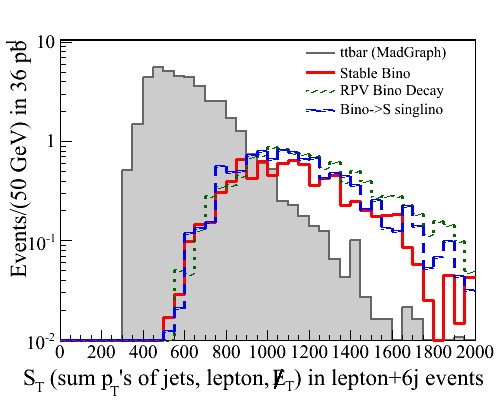}
\caption{\label{basicComparisons} As in Fig.~\ref{basicMET}, distributions for the low-mass benchmark
  signals described in \S \ref{sec:illustrative}.  The \texttt{\texttt{MadGraph}} Monte Carlo estimate of the
  $t\bar t$ background, which dominates for N$_{\text{jets}} \gtrsim 4$, is shown in gray.  
Left: Number of jets ($p_T>30$ GeV) in
  each signal after lepton and $m_T$ requirements.
Right: The $S_T$ distribution [see Eq.~(\ref{STdefn})] 
after requiring 6 jets with $p_T>30$ GeV.  
}
\end{figure}

To demonstrate the impact of reduced  $\MET$ in this case, we show an illustrative
trio of examples in Figures \ref{basicMET} and
\ref{basicComparisons}.  These models will be considered in detail in
\S \ref{sec:illustrative}, to which we refer the reader for a more
complete discussion.  All three share the same
nondescript spectrum of MSSM particles: a 550 GeV gluino, squarks at
800 GeV, and other gauginos obeying approximate mass unification
relations.  The three models differ only in the fate of the bino,
which is taken either to be stable (red solid curves), to decay to a singlet, which decays
to two jets, and a
stable singlino of the NMSSM (green dotted curves), or to decay to three
partons through R-parity violation (blue dashed curves).  Figure~\ref{basicMET} shows the missing energy distribution
for these three models, using base selection cuts similar to those in the CMS jets+$\met$ study~\cite{Collaboration:2011ida}, which require a lepton veto, angular cuts, and $H_T > 300 \GeV$.  The singlino and RPV scenarios have reduced $\met$ and are buried under the background, making them difficult to discover with the standard jets+$\met$ searches.  Figure~\ref{basicComparisons} shows an alternate approach that requires one lepton, $m_T > 20 \GeV$, and no $\met$ requirement.  In this case, the singlino and RPV scenarios are quite distinct from the background in distributions of jet multiplicity and $S_T$.  The lepton-plus-many-jet strategy is therefore complementary to the $\met$-based search.  It provides considerable sensitivity even for the stable bino scenario, and has greatly improved sensitivity to the singlino and RPV scenarios, which would be missed by standard searches.  More details will be given in \S\ref{sec:illustrative}.

\section{Backgrounds to a Lepton-Plus-Many-Jet Sample}
\label{sec:backgrounds}

\begin{figure}[bt]
\includegraphics[width=0.4\textwidth]{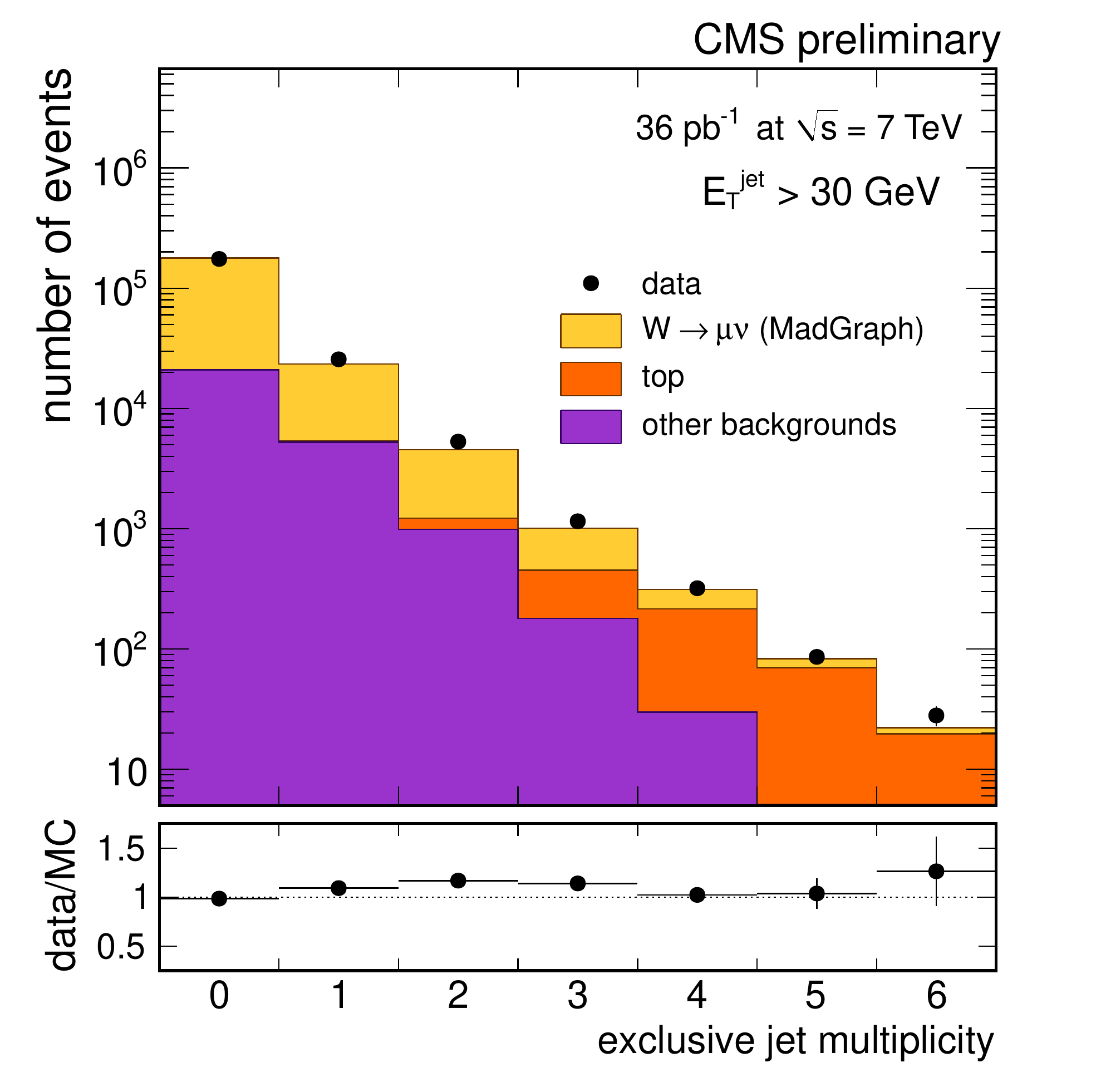}	
\includegraphics[width=0.4\textwidth]{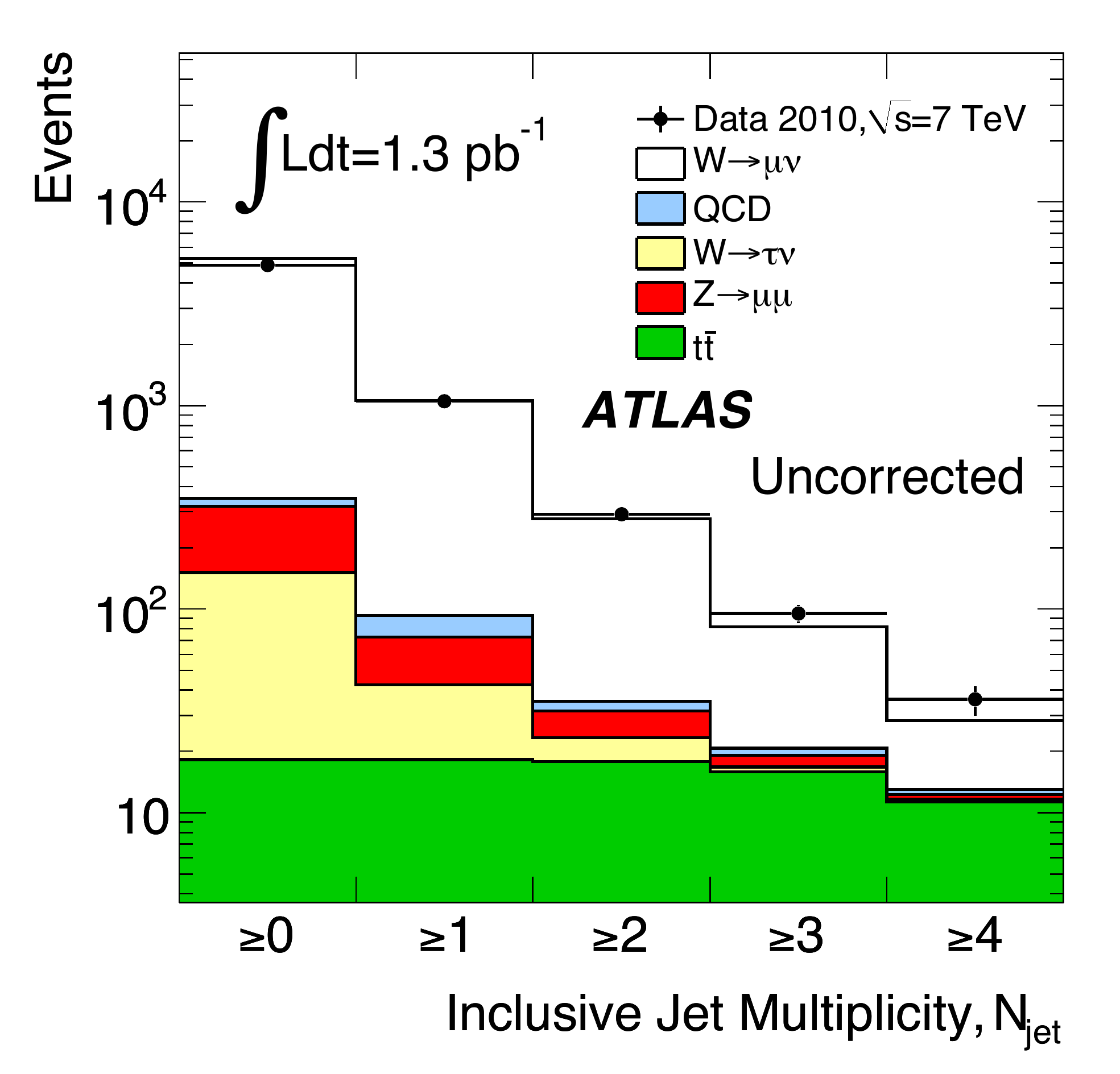}	
\caption{\label{wjetsPlots} 
Distribution of events versus number of jets, taken from $W$-plus-jets studies at CMS \cite{EWKCMS} (left) and ATLAS \cite{EWKATLAS} (right).   Both plots shown are for the muon channel ($p_T^\mu>20$ GeV); plots for electrons are found in the references.  The CMS (ATLAS) plot uses a jet $p_T$ threshold of 30 (20) GeV and requires $m_T>20$ (40) GeV; the  ATLAS plot further requires $\met > 25$ GeV. }
\end{figure}

The relevant backgrounds for a lepton-plus-many-jet search are covered in this section.  A discussion of the Monte Carlo modeling procedure and suggestions for future data-driven background estimates are also included.  More details of the Monte Carlo validation and detector mock-up can be found in Appendix~\ref{app:MCandMockup}.

Production rates for a single lepton plus $n$ jets have been measured at the LHC~\cite{EWKCMS, EWKATLAS}.  Figure~\ref{wjetsPlots} shows the distribution of jet multiplicities in the $\mu$ plus  $n$ jets sample at CMS, for 36 pb$^{-1}$ (left) and at ATLAS for 1.3 pb$^{-1}$ (right).  In both cases, there is good correspondence between the normalization of the data and the Monte Carlo predictions, within statistics.   Note the rate for SM events in the six-jet single lepton channel is \OO(\text{pb}), as seen in the CMS plot in Fig.~\ref{wjetsPlots}. 
At low jet multiplicity, the events are  dominated by $W^{\pm}$ plus $n$ jets, followed by QCD multi-jets where one jet is misidentified as a lepton.  In contrast, the five- and six-jet bins are dominated, for jets of $p_T$ of at least 30 GeV, by $t\bar{t}$ plus jets.  This is because the jets in $W^{\pm}$-plus-jets production are generated through perturbative QCD, and are often forward and/or soft, whereas $t\bar{t}$ production can produce up to four parton-level central jets with sizeable $p_T$, all for the price of $\alpha_s^2$.  

The apparent dominance of the  top background is useful and important, since $t\bar t$ plus one or two jets can be modeled and measured with somewhat more confidence than $W^{\pm}$ plus five or six jets.\footnote{We have checked that $t\bar t$ accompanied by other heavy particles, such as $W$, $Z$, and $h$, are a small contribution.}
Although Fig.~\ref{wjetsPlots} shows that the $W^{\pm}$-plus-jets background is subdominant at high multiplicities, it could potentially become important again on the high $S_T$ tail, where our search is focused.\footnote{We thank G.~Salam for raising this question.}  
At high-$S_T$ the top quarks are increasingly boosted and their daughter jets can merge, or ruin lepton isolation.  This reduces the $t\bar t$ background relative to $W$-plus-jets.  Although we are not aware of evidence that this would make the two backgrounds comparable, this should be checked in data, perhaps using 4- and 5-jet samples.  In addition to the absence of reconstructible top quarks in the $W$-plus-jets background, the rapidity distribution for the lepton is also a handle, as it is central for $t\bar t$-plus-jets (and most signals) and flatter for $W$-plus-jets. (For a recent study out to high multiplicity, see \cite{Stuart}.) 
For the remainder of this work, we assume that the top background remains dominant at the values of $S_T$ (around 1 -- 1.5 TeV) relevant for our studies.  

To study this background, a matched $t\bar{t}$-plus-$n$-jets sample ($n\leq 2$) was generated using \texttt{\texttt{MadGraph} 4.4.49} \cite{Alwall:2007st} for matrix element generation, \texttt{Pythia 6.4} \cite{Sjostrand:2006za} for parton showering and
hadronization, and an MLM matching procedure \cite{mlm} in combination with a shower-$k_{\perp}$ scheme introduced in~\cite{Plehn:2005cq,  Papaefstathiou:2009hp,Alwall:2009zu}.  Further details on the event generation and matching are provided in the Appendix.  The total cross section for the tops is normalized to 150 pb, consistent with the theoretical next-to-leading-order prediction and recent measurements~\cite{TopCMS, TopATLAS}.

The selection cuts used here to study the 36 pb$^{-1}$ samples are modeled after those in the CMS study \cite{EWKCMS}.  (When studying the 1 fb$^{-1}$ samples, we raise these cuts; see \S\ref{ssec:higherLumi}.)  One lepton is required with $p_T > 20 \GeV$ and $|\eta| < 2.1$ for a muon and 2.5 for an electron.  No direct $\MET$ cut is applied; however, it is  required that $m_T = \sqrt{2 p_T \MET (1- \cos\Delta\phi)} > 20 \GeV$.
Jets are formed using an anti-$k_T$ clustering algorithm~\cite{Cacciari:2008gp} in \texttt{FastJet}~\cite{Cacciari:2008gp} with $R=0.5$.  
Jets with $p_T > 30\GeV$ and $|\eta|<2.4$ are counted; those that fall within $\Delta R < 0.2$ of an electron are not included in the jet count.  Our Monte Carlo model has been tuned and validated against data wherever possible (see the Appendix for details).  In particular, the overall rates of three- to six-jet events from $t\bar t$ agree with both data and the CMS full simulation Monte Carlo in Fig.~\ref{wjetsPlots} (see Fig.~\ref{bkgValidation} for comparison).  However, we rely on Monte Carlo for the shape of the $S_T$ distributions and the normalization of the higher-multiplicity jet bins.

\begin{figure}[tbh]
\includegraphics[width=0.45\textwidth]{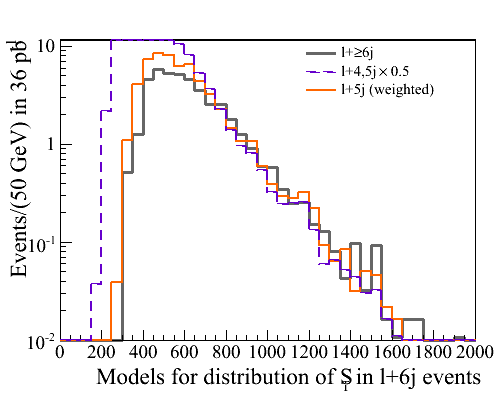}
\includegraphics[width=0.45\textwidth]{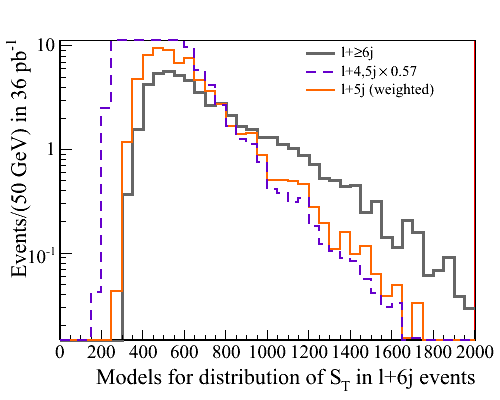}
\caption{\label{fig:STestimators} 
  Suggestion of how lower-multiplicity $S_T$ distributions could be
  useful for a data-driven model of the $t\bar t$ background.  Left:
  The $S_T$ distribution in events with a lepton and $\ge 6$-jet
  events (thick gray), and two distributions from which it could be
  modeled (see text): the $S_T$ distribution of 4 and
  5-jet events (purple dashed) after lepton and $m_T$ selections, and
  5-jet events reweighted  (solid
  orange) as described in Footnote 3.  Right: Same distributions, in the presence of the low-mass
  singlino benchmark [see Eq.~\eqref{benchmarkPoints}], which distorts the
  6-jet $S_T$ distribution but has limited impact at lower
  multiplicity.}
\end{figure}

It should eventually be possible to model the shape of the distribution using data-driven techniques, with little reliance on Monte Carlo.  Preliminary studies in \texttt{MadGraph} and \texttt{Pythia} Monte Carlo suggest that 4- and 5-jet $S_T$ distributions may be a useful tool in estimating the $S_T$ distribution of events with six or more jets.  In particular, the measured power-law of the $S_T$ tails in 4- and 5-jet events can be used to estimate the tail of the higher-multiplicity sample.  To model the full distribution, including its rise and turn-over, a refined method, such as kinematics-dependent reweighting of 5-jet events, is needed.  Both approaches are illustrated in Figure~\ref{fig:STestimators}, in the context of a $t\bar t$+jets Monte Carlo sample (left) and in the presence of a new-physics signal (right).  For background, the lepton plus 4- and 5-jet $S_T$ distribution (dashed purple) models the tail of the 6-jet distribution (thick gray), though it has a much lower threshold and a larger population of low-$S_T$ events.  This difference arises largely from the limited phase-space for lower-$S_T$ events to share this energy among six jets with a fixed $p_T$ threshold of 30 GeV.  A simple reweighting of five-jet events to account for this effect\footnote{Five-jet events are weighted according to a ``splitting factor'' $\sum_{i=1,5} \max\left(0, 1-\frac{2 p_{T,min}}{p_T(j_i)}\right)$, where $p_{T,min}=30$ GeV is the jet $p_T$ threshold.  This function parametrizes the fraction of phase-space in which the splitting of any of the five jets into two lower-$p_T$  jets would produce a six-jet event above threshold.  This naive weight could presumably be improved by QCD-splitting-motivated corrections, and by separately modeling the splitting into 7 or more jets.}
(orange) yields better agreement with the bulk of the $S_T$ distribution.  
A new-physics signal then appears as an excess on the tail of the high-multiplicity $S_T$ distribution relative to the lower-multiplicity curves.

\section{An Illustrative Case Study}\label{sec:illustrative}

This section explores how the lepton-plus-many-jet search strategy can
be applied in a particular case study.  The focus will be on a
simplified model (described in \S\ref{ssec:benchmarks}) within the
MSSM, and two variants thereof in which the lightest MSSM neutralino decays,
decreasing the $\MET$ and increasing the number of jets in a typical event.  \S
\ref{ssec:35pb} considers the lepton-plus-many-jet strategy in the context of the 2010 data 
set of 36 pb$^{-1}$, for which
the existence of published data on $t\bar t$ and other backgrounds
reduces our dependence on Monte Carlo simulations.  Specifically, the
Monte Carlo background generated in this work can be cross-checked
against the published $W$ and $Z$+jets studies at CMS~\cite{EWKCMS}
and ATLAS~\cite{EWKATLAS}.
In this context, evidence is
given that there is complementarity between search strategies based on
missing energy and one based on jet multiplicity and $S_T$.  A
low-mass benchmark point is considered in some detail, and rough
estimates of exclusion reach across the parameter space are given.  

\S
\ref{ssec:higherLumi} then presents prospects for the lepton-plus-many-jet
searches with 1 fb$^{-1}$ of data, accounting for increased $p_T$
thresholds.  A high-mass benchmark point is considered in some detail,
and it is checked that sensitivity to the low-mass benchmark point is retained.

\subsection{The Fiducial Models}
\label{ssec:benchmarks}

We consider a subspace of the MSSM, parametrized by the gluino
pole mass $M_{\tilde g}$ and flavor-universal squark mass $M_{\tilde
  q}$.  The bino and wino soft masses are given by $M_1:M_2:M_{\tilde
  g} = 1:2:6$.  The sleptons and Higgsino are heavy enough that they
play no role.\footnote{Specifically, $\mu = M_{\tilde g} + 100$
  GeV, while slepton masses of 1500 GeV are chosen for convenience.}
This simplified model is loosely reminiscent of the CMSSM and has
similar phenomenology.

The dominant new-physics processes are gluino pair-production, squark pair-production and squark-gluino production.  
When $M_{\tilde q} > M_{\tilde g}$, the gluinos decay predominantly via a two-step cascade
through the wino or Higgsino,
\begin{equation}
\tilde{g}  \rightarrow 2j + \widetilde{W} \rightarrow 2j + W^{\pm}/Z^0 + \widetilde{B}~,
\label{eq: B1}
\end{equation}
and a significant fraction of squarks decay into the gluino.  Therefore,
 it is common to  obtain more than six jets, a lepton from $W$ decay, and (if the bino is stable) missing energy.  
The squarks also have a less active decay:
\begin{equation}
\tilde{q}  \rightarrow j + \widetilde{W} \rightarrow j + W^{\pm}/Z^0 + \widetilde{B}~,
\label{eq: B2}
\end{equation}
which dominates when  $M_{\tilde q}<M_{\tilde g}$.  In this case, high-multiplicity states are not generic unless the bino undergoes further decay.

Three possibilities for the fate of the bino are explored here:
\begin{itemize}
\item A stable bino LSP, giving rise to missing energy.
\item An NMSSM-like decay of the bino into a singlet $S$ and its superpartner, the ``singlino'' $\tilde S$; the $S$ then decays to $b\bar b$. 
($M_S = 0.5 \cdot M_1$, $M_{\tilde S} = 0.25 \cdot M_1$ are taken throughout).  Decays of Higgsino- or wino-like NLSPs to $Z$ or $h$ in gauge mediation \cite{MatchevThomas}, or of any neutral NLSP in a suitable hidden valley \cite{hvsusy}, can have similar kinematics.  In each of these cases, the energy of the NLSP is split between jets and missing energy. 
\item Fully hadronic R-parity violation, in which the bino decays to three quarks.  Only decays to light-flavor quarks will be considered here, though heavy-flavor-rich decays are also possible. The only missing energy in this scenario is from $W$ and $Z$ decays to neutrinos.
\end{itemize}
These three variants of the fiducial model will be compared in the studies presented below.

In addition, two specific points in parameter space will be used to illustrate the effectiveness of the search strategy:
\begin{eqnarray}\label{benchmarkPoints}
\textbf{low-mass benchmark point: }&& (M_{\tilde g}, M_{\tilde q}) = (550, 800) \text{ GeV}   \\  \nonumber
\textbf{high-mass benchmark point: }&& (M_{\tilde g}, M_{\tilde q}) = (700, 1000) \text{ GeV}   \nonumber
\end{eqnarray}

The signal events are generated in \texttt{Pythia 6.4} \cite{Sjostrand:2006za}.  Background simulation is discussed in \S \ref{sec:backgrounds} and Appendix~\ref{appss:ttbackground}.  A detailed description of the detector mock-up can be found in Appendix~\ref{appss:mockup}.  

\subsection{The 2010 data sample of 36 pb$^{-1}$}\label{ssec:35pb}

\begin{figure}[b]
\includegraphics[width=0.45\textwidth]{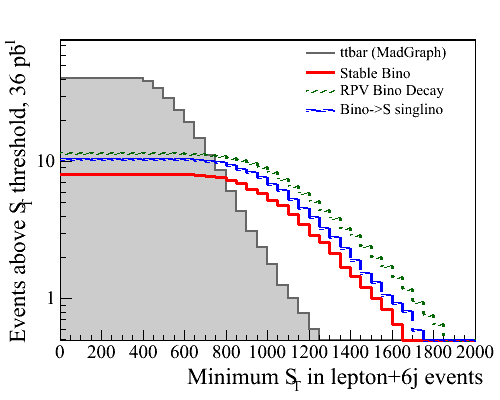}
\includegraphics[width=0.45\textwidth]{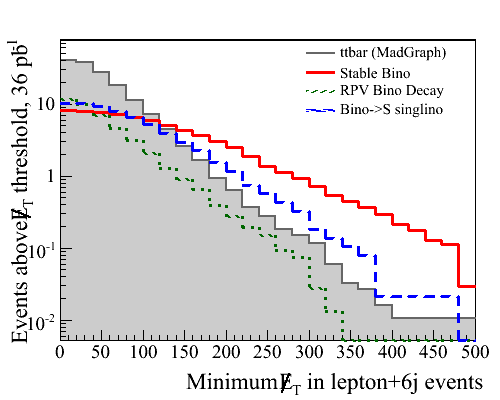}
\caption{\label{STefficiencyPlot} Left: Number of events expected above an $S_T$ cut ({\it i.e.}, the integrated upper tail of the $S_T$ histogram in Fig.~\ref{basicComparisons}), after lepton $p_T$, $m_T$ and 6-jet requirements from \S\ref{sec:backgrounds}.  Right: The integrated missing energy tail after the same selections.  Both plots correspond to the low-mass benchmark.}
\end{figure}
This section explores the effectiveness of the lepton-plus-many-jet search in 36 pb$^{-1}$.  The three variants of the low-mass benchmark --- stable bino, bino$\rightarrow S\tilde S$, and RPV bino decay --- were already considered in Figs.~\ref{basicMET} and \ref{basicComparisons}.  Figure~\ref{STefficiencyPlot} shows the integrated $S_T$ and $\MET$ distributions for the three variants, compared to the matched top background generated with \texttt{\texttt{MadGraph}} and normalized (see \S\ref{sec:backgrounds}) against data \cite{EWKCMS,EWKATLAS}.  For both plots (unlike Fig.~\ref{basicMET}), the $p_T$, $m_T$ and $n_{jets}\geq6$ cuts 
presented in \S \ref{sec:backgrounds} 
are imposed.  While the stable bino scenario produces a small excess on the $\MET$ tail, the two cases with a decaying bino have suppressed missing energy and are not distinguishable from background.  In contrast, all three variants are clearly separated from the background in $S_T$.

\begin{table}[t]
\begin{tabular}{|r|r|r|r|r|}
\hline
 & Stable $\widetilde B$ & $\widetilde B \rightarrow S\ \widetilde S$ & $\widetilde B \rightarrow 3j$ (RPV) & Background \\ 
 \hline
\hline
\textbf{Total Rate (36 pb$^{-1}$)}                         &  86        &  86    &  86   &                              \\ \hline       
\multicolumn{5}{|c|}{\textbf{Lepton+6j}}\\ \hline          
1 lepton ($p_T>20$)                                        &  15        &  15    &  13   &   \emph{1400}                \\        
$m_T(\ell,\nu)>20$                                         &  14        &  12    &  12   &   \emph{1200}                \\        
$\geq 4$ jets ($p_T>30$)                                   &  13        &  12    &  12   &   \emph{460}                 \\         
$\geq 6$ jets ($p_T>30$)                                   &  7.5       &  9.7   &  11   &   \emph{43}                  \\        
$S_T> 1000$ GeV                                            &  4.9       &  6.3   &  7.7  &   \emph{2.2}                 \\ \hline  
\multicolumn{5}{|c|}{\textbf{CMS high-$H_T$}}\\ \hline     
Lepton veto                                                &  70        &  71    &  73   &                              \\
$\geq 3$ jets ($p_T>50$)                                   &  69        &  71    &  73   &                              \\
$d\phi(j,\MET)$                                            &  54        &  51    &  51   &                              \\
$H_T > 500$ GeV                                            &  50        &  50    &  50   &                              \\
$\MET > 150$ GeV                                           &  25        &  9.1   &  1.8  & $\mathbf{43.8 \pm 9.2}$      \\ \hline
\multicolumn{5}{|c|}{\textbf{ATLAS jets+MET D}}\\ \hline   
3 jets, $p_T>$120, 40, 40                                  &  85        &  86    &  86   &                              \\
Lepton veto                                                &  64        &  65    &  68   &                              \\
$\MET>100$                                                 &  47        &  26    &  7.5  &                              \\
$d\phi(j,\MET)$                                            &  33        &  16    &  4    &                              \\
$\MET/M_{eff} > 0.25$                                      &  15        &  3.1   &  0.3  &                              \\
$M_{eff}>1000$ GeV                                         &  3.1       &  0.4   &  0.04 &  $\mathbf{2.5 \pm 1.0_{-0.6}^{+1.0}  \pm 0.2 }$ \\
\hline

\end{tabular}
\caption{\label{tab: cuteff}Total expected event rate in 36 pb$^{-1}$ for the low-mass benchmark models discussed in \S \ref{ssec:35pb}, and the rates surviving several cut flows in Monte Carlo.  The top section of the table displays the cut-flow efficiencies for the lepton-plus-many-jet strategy, with background rates in italics corresponding to $t\bar t$ only, as computed using \texttt{MadGraph} Monte Carlo.  The lower two sections give estimated efficiencies for the same models, through cut-flows based on the hadronic jets+$\MET$ analyses done by CMS (the high-$H_T$ selection used in \cite{Collaboration:2011ida}) and ATLAS (selection `D' used in \cite{daCosta:2011qk}). For the latter two cases, the (bold-faced) expected background is  the full background prediction quoted in the respective search papers.}
\end{table}
Table~\ref{tab: cuteff} summarizes the estimated efficiencies at each stage of the lepton-plus-many-jet selection for each variant of the low-mass benchmark point.  After imposing $p_T$ requirements on the lepton and jets, a multiplicity requirement of six or more jets, and an $S_T$ cut of 1000 GeV, there are 4.9, 6.3, and 7.7 signal events remaining for the stable $\widetilde{B}$, $\widetilde{B} \rightarrow S \widetilde{S}$, and $\widetilde{B}$ RPV decay variants, respectively, compared to 2.2 events for the top background.  
Even with no assumptions about the top background,  the models with decaying binos can be easily excluded if only the expected 2 events are observed.  For models where the excess over background is not as significant, a robust understanding of the top background must be obtained through a reliable data-driven model (or a hybrid Monte-Carlo-and-data-driven model).  Issues relevant to this step are discussed in \S\ref{sec:backgrounds}; see also \S\ref{sec:commentsOnCrossChecks}.

Table~\ref{tab: cuteff}  also summarizes the cut-efficiencies for two high-$\MET$ selections modeled after
jets+$\MET$ analyses recently published by CMS (the high-$H_T$ selection from \cite{Collaboration:2011ida}) and ATLAS (selection `D' from \cite{daCosta:2011qk}).  (Though we have attempted to mimic the selections in these references, we emphasize that our detector mock-up (described in Appendix \ref{appss:mockup}) is highly simplified, and the results are only approximate.)
Even with a stable LSP, the lepton-plus-many-jet search has sensitivity only slightly weaker than the existing
high-$\MET$ searches.  And the lepton-plus-many-jet search gains sensitivity if the LSP decays, while the other search strategies rapidly lose their power.

\begin{figure}[tb]
\includegraphics[width=0.95\textwidth]{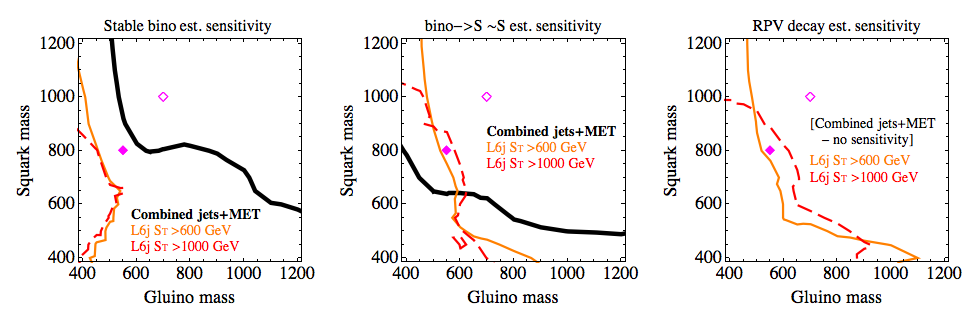}
\caption{\label{fig:parameterScan} Estimated 36 pb$^{-1}$ sensitivity of lepton-plus-many-jet searches, with $S_T$ thresholds of 600 GeV (orange) and 1000 GeV (red dashed), for the fiducial models as a function of different squark and gluino masses.  Each panel corresponds to a different LSP scenario: stable $\tilde B$ (left),  $\tilde B \rightarrow S + \tilde S$ (middle), and hadronic RPV decay of $\tilde B$ (right).  
For comparison, we estimate the combined sensitivity of three $\approx 36$ pb$^{-1}$ hadronic SUSY searches in ATLAS \cite{daCosta:2011qk} (region ``D") and CMS \cite{Collaboration:2011ida} (high-$H_T$ and high-$\mht$)  (thick black).  The low- and high-mass benchmark points are marked by solid and open magenta diamonds.}
\end{figure}

Figure~\ref{fig:parameterScan} generalizes these results to a broad range of gluino and squark masses.
Rough estimates of the expected sensitivity of the lepton-plus-many-jet and existing
high-$\MET$ searches at 36 pb$^{-1}$ are shown as a function of $M_{\tilde{g}}$ and $M_{\tilde q}$ for the stable
bino (left), bino $\rightarrow$ singlino (center), and RPV (right) variations.
The orange solid (red dashed) line is the projection of the likely limits
for the lepton-plus-many-jet search with $S_T > 600\ (1000)\ \GeV$. For the 600 (1000) GeV cut, the
expected top background is 15 (2) events, and it is assumed that a signal of 12 (6)
expected new-physics events passing the selection could be excluded.  
This exclusion threshold assumes that systematic errors on the $t\bar
  t$ background are small enough to be neglected compared to
  statistical errors.  A significant systematic error would reduce the overall sensitivity
  of a lepton-plus-many-jet search, but would not alter our conclusion that the search is complementary to
  missing-energy-based searches.

For comparison, the estimated reach of jets+$\met$ searches for these models is also shown in Fig.~\ref{fig:parameterScan}. The thick black lines on the plots are a rough estimate of a combined 
limit from the ATLAS and CMS jets-plus-$\MET$ searches, in particular \cite{daCosta:2011qk} (region ``D") 
and CMS \cite{Collaboration:2011ida} (high-$H_T$ and high-$\mht$ regions).  These estimates neglect any signal contamination
of control regions, which could weaken the projections by up to $\sim$50 GeV.  

The combination of jets+$\MET$ analyses gives the best
coverage for the stable bino scenario, where the mass splitting between the gluino and bino is large
and there is considerable missing energy.  When the bino decays to a singlino, the jets+$\MET$ searches lose
their reach; fewer events survive the strong $\MET$ cut, and limits can only be set when the gluino and
squark masses are light enough to maintain a large production
rate.  In contrast, the lepton-plus-many-jet analysis has significant coverage for gluino
masses less than $600$ GeV and all squark masses up to $1200$ GeV, and 
can reach up to $m_{\tilde g}\sim 900$ GeV for lower squark masses.
With an RPV decay of the LSP, the missing energy arises only from the $W^{\pm}$ that produces the
lepton, and the jets+$\MET$ search has essentially no sensitivity.  The lepton-plus-many-jet analysis largely compensates for this loss.

B-tagging and anti-tagging are important handles for both classes of searches.  Two ATLAS searches requiring b-tags~\cite{Aad:2011ks}, not shown, may extend the light-squark sensitivity for the singlino model by $\sim 50$ GeV.  The use of $b$-tagging fractions in a lepton-plus-many-jet search is discussed in  \S \ref{ssec:tagging}.

\subsection{Prospects for Searches in 2011 Data}\label{ssec:higherLumi}

In this section, the potential sensitivity of  lepton-plus-many-jet searches in 1 fb$^{-1}$ of 2011 data is briefly
considered.  The complementarity between lepton-plus-many-jet and jets+$\MET$ searches, illustrated at 36 pb$^{-1}$ in \S \ref{ssec:35pb}, should persist at higher luminosities, but as there is as yet no data with which to normalize Monte Carlo estimates, this cannot be studied reliably. The discussion here is therefore limited to a semi-quantitative and preliminary examination of the two benchmark points.

LHC data-taking during the ongoing 2011 run is at much higher
luminosities than in 2010, leading to higher trigger thresholds (on
$H_T$, jet and lepton $p_T$, etc.)  and significant pile-up.
Consequently, a lepton-plus-many-jet search in 2011 data will
presumably require tighter cuts than those used in \S \ref{ssec:35pb}.
The following discussion therefore proceeds with 
lepton $p_T$ and $m_T$ requirements  raised from 20 up to 30 GeV,
and with jet $p_T$ threshold raised from 30 to 45 GeV.

\begin{figure}[b]
\includegraphics[width=0.45\textwidth]{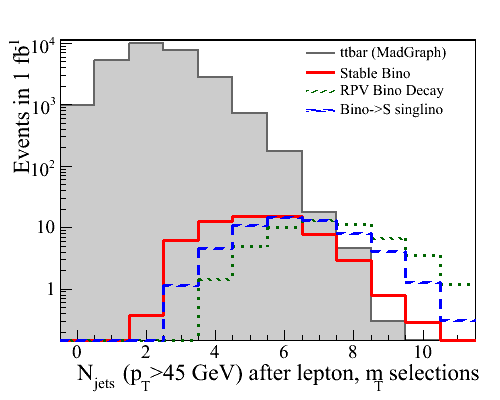}
\includegraphics[width=0.45\textwidth]{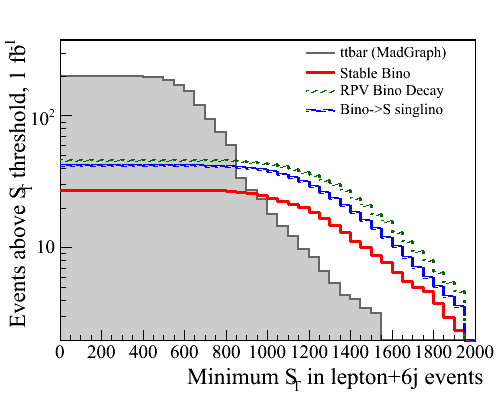}
\caption{\label{HLumiSignalsPlots} 
Kinematic distributions in the signal region, for the {\it high-mass} benchmark models, with the tighter cuts (jet $p_T > 45$ GeV) 
and 1 fb$^{-1}$ integrated luminosity: (left) $n_{jet}(p_T \geq 45 \mbox{ GeV})$ after lepton and $m_T$ cuts and (right) integrated number of events above an $S_T$ cut, after additionally requiring at least 6 jets.
}
\end{figure}
Despite the higher thresholds, the separation of signal from
background \emph{improves} relative to what is seen in Table \ref{tab: cuteff}.
Under the new cuts, combined with a raised $S_T$ cut of 1000 GeV, the
efficiency for the top background is reduced by a factor of 3, while
the efficiency of the low-mass benchmark signals, for all three
variants, drops by no more than a factor of 2.  
The much greater integrated luminosity of 1 fb$^{-1}$ easily compensates for the lower signal efficiency,
and allows use of seven-jet events, as well as cross-checks in other distributions (see \S\ref{sec:commentsOnCrossChecks}).  Thus, higher luminosity will enhance sensitivity to lower-mass scenarios, despite the increased thresholds.

Of course, higher luminosity also allows the use of the lepton-plus-many-jet approach for much heavier gluino and squark masses than the low-mass benchmark.  
Figure~\ref{HLumiSignalsPlots} shows multiplicity and kinematic distributions
for the three variants of the
high-mass benchmark point, with $(M_{\tilde g}, M_{\tilde q}) = (700, 1000) \text{ GeV} $.  
With a decaying bino,
the shapes of the signal-plus-background
distributions, both in $n_{jet}$ after an $S_T$ cut and in $S_T$
after an $n_{jet}$ cut, are quite different from the top background.
Furthermore,
with the larger gluino and squark masses, the signal $S_T$
distribution peaks at higher values, and is thus more easily separated
from the $t\bar t$ background.

\section{Potential Cross-Checks on Backgrounds and Signals}\label{sec:commentsOnCrossChecks}

In this short section, we consider several variables that could 
distinguish a new signal from a mismodeled background.  Among
these, the most powerful (though not for all signals) appears to be
the $b$-tag multiplicity distribution, discussed in
\S\ref{ssec:tagging}.  A number of other
variables that are also worth considering are briefly mentioned in \S\ref{ssec:otherCrossChecks}.

\subsection{Cross-checks using heavy-flavor-tagged jet multiplicities}
\label{ssec:tagging}

Top quark samples have well-measured and well-understood $b$-tagging
rates, arising from the ubiquitous $b$ and $\bar b$ and the presence of a $c$ quark
in half the mixed leptonic-hadronic events.  
A distortion in the expected
ratios of tag-multiplicities is evidence for a signal.  Of course,
some signals, such as those that typically have a $t\bar t$ or $b\bar
b$ pair in every event, have similar $b$-content to background and
 show little distortion.  In this case, other variables, such as
those mentioned in \S \ref{ssec:otherCrossChecks}, may be needed.

\begin{figure}[htbp!]
\includegraphics[width=0.45\textwidth]{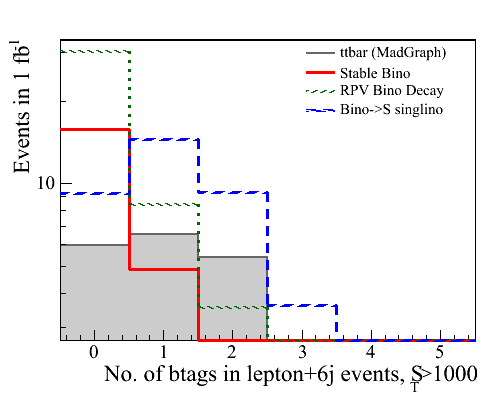}
\caption{\label{btagCorrelation} 
Estimate of $b$-tag multiplicities for \texttt{MadGraph} $t\bar t$ (gray shaded, not including $c$ tags) and the three high-mass benchmark models: (red solid) stable LSP, (green dotted) LSP decay through singlino and singlet, (blue dashed) RPV decay of LSP.  We assume
 `naive' tagging fractions, as described in the text. }
\end{figure}

Estimates of the $b$-tagging multiplicity distributions for the three
variants of our high-mass benchmark model are shown in
Fig.~\ref{btagCorrelation}, for 1 fb$^{-1}$ integrated luminosity and a minimum jet $p_T$ of 45 GeV.  A
similar estimate for the top background is shown.  These numbers are obtained
very naively, by requiring that the $b$ jet be in the tracking
volume and assuming a $60\%$ tagging efficiency.  Charm tags and
mistags are not accounted for, but should not be important for signal.
For background, results such as Figures 1d and 2d in the auxiliary plots 
of \cite{ATLAS1lepton} suggest that charm tagging 
would contribute of order one event to the three-tag bin.
Other backgrounds such as $t\bar t b\bar b$, $t\bar t h$, etc., appear
negligible.  The different variant models exhibit significant differences in $b$-tagging
multiplicities.  The stable-LSP and RPV models, which mainly produce
 light quark jets, give a large contribution to the $0$-tag bin
relative to the $1$-tag bin.  The singlino model, with its four extra $b$ quarks
per event, has a significant number of events with more than one $b$ tag.  In the following section,
we will see other examples of models that give even larger excesses in the
3-tag bin.
Clearly, distributions of $b$-tag multiplicities can be very
helpful in separating an excess from signal from a mismodeled top
background.

One challenge for this method is that at high $S_T$ some of the jets
can be quite hard, and tagging fractions become both smaller and less
precisely known.  Fortunately it may be possible to use data to determine
the tagging fractions in the top background.  The extra jets in $t\bar
t$ plus two jets are rarely $b$ or $c$ jets, and so the tagging
fractions in $t\bar t$ plus zero, one and two jets may be quite
similar.  However, this suggestion has not been checked in Monte Carlo.

\subsection{Other Signal Cross-Checks}
\label{ssec:otherCrossChecks}

There are a number of other features that can make a sample with signal-plus-background look different from a background that is mainly from $t\bar t$-plus-jets.
These include the following:
\begin{itemize}
\item The jet multiplicity distribution, which for $n_{jets}\ge 6$
  falls off gradually in the top background, but can be nearly flat in
  the presence of signal from 6 out to 7 and 8 jets.
\item The $\eta$ distribution, event-by-event, of the leading six
  jets.  In signal, the jets are more tightly clustered around their average
  $\eta$ than in background, where one or two of the jets
  tend to be from ISR.
\item The total integrated lepton-charge asymmetry, which is small in $t\bar t$ (perhaps of order a few percent)
but can be as large as $2:1$ in some signals, for instance one dominated by associated squark-gluino production.
\item The efficiency for reconstructing top quarks, which can be greatly reduced for signals in which top quarks
are rare.
\item Angular variables that are characteristically correlated with one another in top quark backgrounds.  For instance, the observables $\Delta R$ between the two leading jets, $\Delta R$ between the leading jet and the lepton, and $\Delta \phi$ between the lepton and the $\MET$ vector show strong correlations in background.  These correlations
are often reduced  in signals.  
\end{itemize}
Note that some of these variables, in particular the first and second, are also useful in rejecting $W$-plus-jets backgrounds.

For our high-mass benchmarks, 1 fb$^{-1}$ may be too small for these
features to be significant individually, though the first two seem
promising even at rather low statistics. With several times more data, 
all of these observables should become useful.
Combining these variables might also prove powerful,
though this has not been studied here.

\section{Examples of Other Models}
\label{sec:otherModels}

Having considered a particular class of fiducial models in detail, we
now check that the lepton-plus-many-jet search is
sensitive to a wide range of signals.  We have studied a number of
different classes of models.  For some classes, the results are
similar to those in the fiducial models, while for others some new
features arise, particularly in the context of $b$-tagging.

\subsection{$t$-rich SUSY Models}

Many SUSY models are a large source for top quarks, which naturally
result in high-jet-multiplicity signals.  We have studied one class of examples,
with relatively light stops and sbottoms, a bino LSP, and a gluino
somewhat heavier than the stops and sbottoms but lighter than the
other squarks.  Although such models often have large cross-sections
for electroweak gaugino production, much of the relevant cross-section
for high-multiplicity events comes from associated squark-gluino
production, with gluino and squark pair production also playing a
role.  As before, we have chosen three variants of these models, with
a stable bino, a bino decaying to a singlet and singlino, and a bino decaying
to light quarks via R-parity violation.

Almost all of our plots (omitted for brevity) are qualitatively the
same as those for the fiducial models shown in \S\ref{ssec:35pb} and 
\S\ref{ssec:higherLumi}, and our conclusions remain the same.
The one exception is in $b$-tagging, which, because of the presence of
top and bottom quarks in many of the events, is shifted to larger tag
multiplicity relative to the models shown in
Fig.~\ref{btagCorrelation}.  In the following, more examples of theories
with an excess of $\geq$ 3 tagged jets will be given.

\subsection{GMSB with $Z$ or $H$ decays}
\label{ssec:GMSB}

In gauge-mediated SUSY breaking (GMSB), the lightest
SM superpartner is the NLSP and decays to a gravitino, the true
LSP.  Such models can also give high-multiplicity signals with reduced
$\MET$.  Here, we consider the case of a Higgsino NLSP
decaying \cite{MatchevThomas} mainly to a $Z$ or $h$ plus a gravitino,
with a very small branching fraction to decay to a photon plus
gravitino.  Similar models, possibly with even higher
multiplicity, can arise in the presence of a Hidden Valley sector
\cite{hvsusy}.
\begin{figure}[tb]
\includegraphics[width=0.45\textwidth]{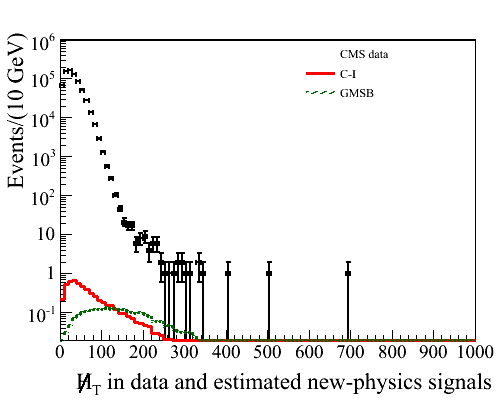}
\includegraphics[width=0.45\textwidth]{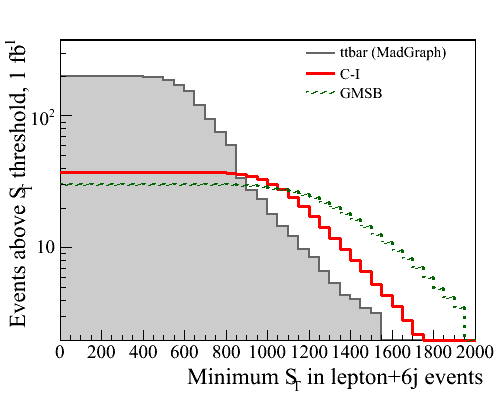}
\caption{
\label{otherMETST} 
Distributions 
for the gauge-mediated model (green dashed) and the compositeness-inspired model (red solid) discussed
in \S \ref{ssec:GMSB} and \S \ref{ssec:composite} respectively.
Left: Missing energy distributions, as in Fig.~\ref{basicMET} (36 pb$^{-1}$).     
Right: Number of events expected above an $S_T$ cut (i.e. integrated upper tail of $S_T$ histogram) after lepton $p_T$, $m_T$ and 6-jet requirements, as in Fig.~\ref{HLumiSignalsPlots} (1 fb$^{-1}$).  
}\end{figure}

As an illustration, we choose a
gauge-mediated model with a Higgs boson at 120 GeV and Higgsinos
$\tilde h$ at 198--208 GeV.  The Higgsino NLSP has branching
fractions of $78\%$, $20\%$ and $2\%$ to $h$, $Z$ and $\gamma$ plus a
gravitino.  We consider a model in the class of General Gauge
Mediation (or, more simply, with a non-minimal set of messenger fields
for which colored and colorless messengers are not mass-degenerate).
In this model, the gluino is at 800 GeV, the bino at 404 GeV and the
wino at 856 GeV; the squarks are at 1100 GeV.  Sleptons are near 500 GeV,
too heavy to play any role.  Dominant production modes
involve gluino pairs, with $\tilde g\to t\bar b \tilde h^-, \bar t b
\tilde h^+, \bar t\bar t\tilde h^0$, with a minority of gluinos decaying
to light quark pairs plus a Higgsino.  The large number of $b$ quarks
in the final state is further enhanced when the Higgsino decays to a
Higgs boson.

The model produces numerous same-sign di-lepton events, but we estimate that the
number of events in 1 fb$^{-1}$ that pass the stringent
requirements of the 2010 CMS search \cite{Chatrchyan:2011wb} is still
rather small, making our current search at worst complementary.  In
fact, because trigger and analysis thresholds for the same-sign search
must presumably be raised for 2011, our search strategy might prove
more sensitive.

\begin{figure}[t]
\includegraphics[width=0.45\textwidth]{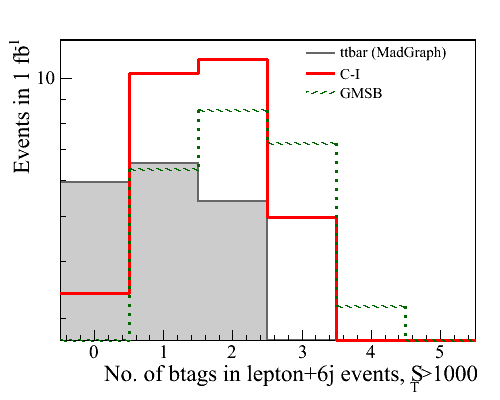}
\caption{\label{btagCorrelation2} 
As in Fig.~\ref{btagCorrelation}, estimated $b$-tag multiplicities at 1 fb$^{-1}$ for \texttt{MadGraph} $t\bar t$ (gray shaded, not including charm tags) and 
for the gauge-mediated model (green dashed) and the compositeness-inspired model (red solid) discussed
in \S \ref{ssec:GMSB} and \S \ref{ssec:composite} respectively.}
\end{figure}

The integrated $S_T$ distribution for this model is presented in Fig.~\ref{otherMETST} and demonstrates that this
model well exceeds the top background using our methods.  The
$\MET$ distribution, also shown in Fig.~\ref{otherMETST} (where the background, as in Fig.~\ref{basicMET},
is the 
data taken from \cite{Collaboration:2011ida}), confirms that this model is not
easily seen using the simplest jet-plus-$\MET$ searches.  Meanwhile, the $b$-tag distribution, shown
in Fig.~\ref{btagCorrelation2}, demonstrates that the excess is quite different from the expected 
top background, with over a third 
of the excess events carrying three or more $b$ tags.

\subsection{Non-SUSY models with top quarks plus other particles}
\label{ssec:composite}

Certain strong-dynamics BSM models \cite{compositeness} may contain new composite fermions $\psi$ in
higher-representations of $SU(3)$, such as sextets or octets of color.
Other models may have multiple triplets. 
These fermions may couple
most strongly to the third generation, and may largely decay to a top
quark plus an exotic scalar or pseudoscalar $\phi$.  The field
$\phi$ may be an octet (triplet) [singlet] of color if $\psi$ is a sextet or triplet (octet) [triplet].
This scalar may then in turn decay to gluons or to heavy-quark pairs.
Alternatively, this scalar may be too heavy to be produced on-shell, in which case
the decay of $\psi$ may be a three-body decay to a top quark plus two other hard partons.

Another class of models with similar phenomenology can arise in
the context of R-parity-violating SUSY.  A gluino (playing
the role of $\psi$) may decay to the top squark $\tilde t$ (playing
the role of $\phi$) which then decays $\tilde t\to\bar b \bar s$.  
Alternatively, the gluino may decay directly to a triplet $tbs$ of
quarks, as a three-body decay.

For technical reasons, we have chosen to display the RPV SUSY model
with $\tilde g\to \tilde t \bar t$, followed by $\tilde t\to bs$, though again
we emphasize that there is nothing especially supersymmetric about this
signature.  The other models are almost identical in most kinematic distributions,
except for normalization, on which we comment below.  The
integrated $S_T$ distribution, and the $\MET$ distribution, are shown
in Fig.~\ref{otherMETST}, while the $b$-tagged
jet multiplicity for this particular model (with 4 $b$ quarks in each
event) is shown in Fig.~\ref{btagCorrelation2}.

In comparing the particular model simulated here to the non-SUSY compositeness-inspired
models mentioned in the first paragraph of this section, there are a few things to
keep in mind.  
First, Dirac octet fermions, as complex representations, have double
the cross-section of a Majorana gluino at the same mass scale. Second,
sextet fermions and Dirac octet fermions have almost identical
cross-sections and kinematics in their decays.  Third, although
Majorana gluinos will also produce a same-sign lepton signal, sextets
or Dirac octets, being complex, will not.  Thus, while the Majorana
gluino case shown in the figures is somewhat borderline for detection
using our method, the sextet and Dirac octet would be more easily
detectable, and invisible to the same-sign di-lepton search.

Meanwhile, triplets, such as a $t'$, have a cross-section several
times smaller than the gluino.  Our methods would need further
optimization to detect one such quark, for a wide range of masses.
However, a model with
several roughly-degenerate quarks should be observable using our
proposed analysis.

Finally, while the RPV SUSY model in our plots produces four bottom
quarks in each event, which will further distinguish it from
backgrounds, the non-SUSY models mentioned above might
produce as few as two and as many as six, depending on the decay mode
of the scalar $\phi$.  Consequently, $b$-tagging is very useful as a
cross-check in some models of this type, though not in all.

\subsection{Departures from Perturbative Quantum Field Theory}

The lepton-plus-many-jet strategy is quite inclusive, so it is also potentially sensitive to new physics that is not well understood theoretically.
For example, a breakdown of perturbative quantum field theory that
leads to large partonic cross-sections would potentially be detectable
through such a search.  In particular, models that have partonic
cross-sections that rise at high $S_T$, but less dramatically than
black holes, might not be detectable without requring a lepton in
order to reduce QCD backgrounds (see Fig.~\ref{bhPlots}).  While we
have not studied specific models, it appears that a
lepton-plus-many-jet search would fill a gap between standard black
hole searches and other exotica and SUSY searches.

\section{Discussion and Conclusions}\label{sec:conclusion}

In this work, we have argued that a search requiring high jet
multiplicity and high $S_T$, along with a lepton, and with a limited
$m_T$ requirement (for the purpose of reducing fake leptons),
has sensitivity to phenomena that are currently not being covered by
existing $\MET$-based SUSY searches or by any published exotica
searches.  A key observation is that the background is apparently
dominated by top-quark backgrounds, not by $W$-plus-jets. Checking
our work where possible against existing public results,
we have argued (see Fig.~\ref{fig:parameterScan}) that this
strategy, applied to the 2010 data, would be sensitive to large classes
of models that are not excluded by the existing searches.  We have
further argued that this strategy still works with cuts raised
somewhat to account for higher trigger thresholds and pile-up in the
2011 data.  Furthermore, this search is sensitive to many different
types of high-cross-section physics, including a large variety of SUSY
models and compositeness models.  We hope that searches along these
lines are already underway, or will soon be undertaken.

Before focusing on the lepton-plus-many-jet sample, we argued more
generally that strategies based on high-multiplicity, high-$S_T$, and
a non-jet visible object could be powerful.  It is natural to consider replacing
the lepton in our search with a photon or $Z$, or to consider
something a bit more elaborate, such as same-flavor opposite-sign
leptons off the $Z$ peak (non-$Z$ di-leptons, or ``NZDL''.)  The NZDL
case seems most likely to extend directly from our
current discussion.  In some kinematic regimes, 
the background
is again dominated by top, and is relatively simple to model.  For signals
that dominantly produce leptons in pairs, through an off-shell $Z$ or
in a cascade decay with a slepton-like intermediate state, a strategy
requiring an NZDL pair and several jets (perhaps as few as four) might
do better than the single-lepton-plus-six-jet strategy we have described in this
paper.

Searches involving a $Z$ boson, on the other hand, may have either
$Z$-plus-many-jet or $t\bar t$ as a background, depending on cuts.  In
the latter case, the extra $Z$ bosons themselves are a signal; in the
former case, however, a strategy such as ours might apply.  The
challenge here is to determine the background, perhaps using
photon-plus-many-jet to estimate the $Z$-plus-many-jet distributions.

For a search in the photon-plus-many-jet channel itself, however, background
modeling may be difficult, as there is no obvious data-driven method.
The strategy suggested in \cite{Stuart}, of looking at high
jet-multiplicity for an excess of photons or $Z$ bosons at central
rapidity, would apply, but it would be interesting to seek an
alternate approach.

More generally, we strongly encourage the experimental groups,
whenever practical, to make public the $S_T$ distributions for their
samples, including Standard Model measurements and control regions.
Although we recognize that validating such plots is by no means simple, they are of
great physical interest.  Among the various possible
kinematic variables, it appears that $S_T$ is especially robust and
informative.  Distributions in $S_T$ would assist
theorists in confirming their estimates of backgrounds, and estimating whether high-mass
signals of interest are within reach of existing data.  It may even be possible
to exclude some models with large cross-sections, based only on the overall high-$S_T$ rate without
a dedicated search. 

Clearly, a diverse array of LHC searches is necessary so that new physics can be found wherever it may be, and to assure maximal use is made of LHC data.
To develop a sufficiently broad search program requires identifying many regions of phase-space where Standard Model backgrounds
are small and plausible new-physics signals could be present at detectable levels.  We have studied one such region (and suggested a few others) in this paper, and argued that reasonable signals may be large where the backgrounds are small.  Meanwhile, if one is to draw
any general scientific lesson (such as the absence of superpartners below some mass scale)  from a collection of null results,
it is crucial that this set of searches be robust to modifications of the theory in question.  We have argued that combining the lepton-plus-many-jet search (along with other similar searches) with existing $\MET$-based searches significantly improves the robustness of the set.

For any study of this type, Standard Model backgrounds are the central concern, and thus careful measurements of the Standard Model at the LHC are the key to the effort.  This is  especially true when the leading backgrounds
are difficult to calculate reliably.   For us, the $W$- and $Z$-plus-jets measurements at ATLAS and CMS \cite{EWKATLAS,EWKCMS}
were invaluable, both in motivating our study of the lepton-plus-many-jets strategy and in allowing us to
think through certain subtleties that such an analysis would encounter.   We expect that the growing body of LHC Standard Model measurements will suggest many new avenues for future searches.

\appendix
\section{Monte Carlo and Detector Mock-Up}\label{app:MCandMockup}
\subsection{Modeling the $t\bar t$ plus jets background}\label{appss:ttbackground}

\begin{figure}[tb] 
   \centering
   \includegraphics[width=2.5in]{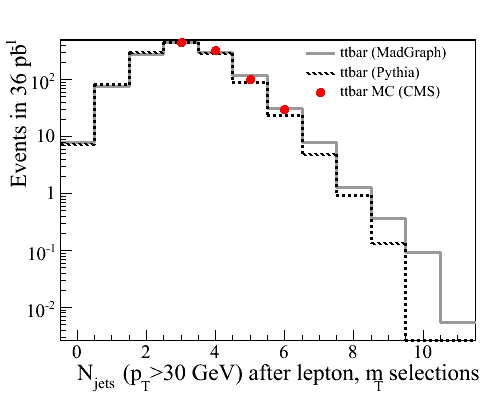} 
   \includegraphics[width=2.5in]{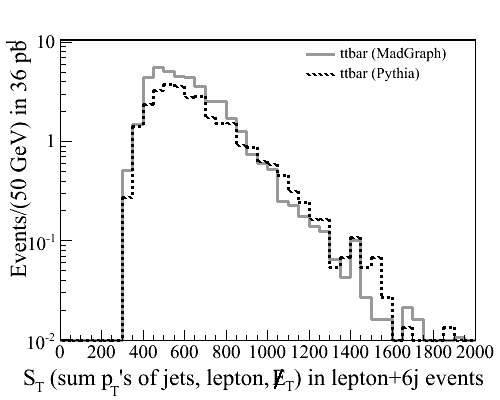}
   \caption{ Left: Our computation of the jet multiplicity distribution for matched \texttt{MadGraph} $t\bar{t}$ (gray) and Pythia $t\bar{t}$ (black dotted) Monte Carlo samples, for electrons and muons combined.  The red dots indicate the CMS Monte Carlo (consistent with 36 pb$^{-1}$ data) in each bin, as shown in Fig.~\ref{wjetsPlots}.  Right: Comparison of $S_T$ distributions for the matched \texttt{MadGraph} and unmatched \texttt{Pythia} $t\bar{t}$ samples.}
   \label{bkgValidation}
\end{figure}

A matched $t\bar{t}$ background was generated with \texttt{\texttt{MadGraph}
  4.4.49} \cite{Alwall:2007st} with CTEQ6L1 parton distributions
\cite{Pumplin:2002vw}.  To guarantee that the high $S_T$ tail of the
top distribution is well-populated, 50K events were generated in each
of three regimes, where the top with the highest $p_T$ in the event
was required to either be less than 100 $\GeV$, between $100$ - $300
\GeV$, or greater than 300$\GeV$.  \texttt{Pythia 6.4.22} 
\cite{Sjostrand:2006za} was used for parton showering and
hadronization.  An MLM matching procedure was implemented using
\texttt{\texttt{MadGraph}} and \texttt{Pythia} in combination with a
shower-$k_{\perp}$ scheme introduced in~\cite{Plehn:2005cq,
  Papaefstathiou:2009hp,Alwall:2009zu}.  The matching scale was
$Q_{\text{Match}} = 100$ GeV for the top sample.  The results were then passed through a private detector mock-up as described below.

Figure \ref{bkgValidation} (left) compares our matched $t\bar{t}$ sample with  the Monte Carlo expectation given by CMS in \cite{EWKCMS}.  Our sample is normalized to the top NLO cross-section of 150 pb, giving a K factor of 1.64.  The three and four jet channels agree perfectly, but our Monte Carlo over-predicts the five and six-jet channel by approximately 20 and 30\%, respectively.

We also generated an unmatched top sample using \texttt{Pythia}. Figure~\ref{bkgValidation} (right) shows the $S_T$ distributions for the \texttt{Pythia} and \texttt{MadGraph} samples.  $\texttt{\texttt{MadGraph}}$ with matching gives more six-jet events than does \texttt{Pythia}; this is expected, as the matching procedure includes two additional ISR jets in the matrix element, as compared to \texttt{Pythia}, which only includes one by default.  The $p_T$ distributions for the hardest four jets in each Monte Carlo sample correspond well with each other, leading to good agreement on the high-$S_T$ tail in~Fig.~\ref{bkgValidation}.  Meanwhile, although the $p_T$ distributions of the fifth and sixth jet differ between the matched and unmatched samples, this has a significant effect only on the low-$S_T$ bins of Fig.~\ref{bkgValidation}.

\subsection{Reconstruction and Analysis/Detector Mock-Up}\label{appss:mockup}

We employ a simple detector mock-up that allows flexibility in tuning the simulated 
detector response and reconstruction efficiency for any given analysis.  

Starting from hadronized Monte Carlo truth, we build jet objects with \texttt{FastJet} \cite{Cacciari:2008gp}, using anti-$k_T$ with $\Delta R = 0.5$. 
All hadrons with $|\eta|<3.0$ and leptons (electrons and muons) and photons with $|\eta| <2.5$ are included in the jet reconstruction.  A two-dimensional missing energy vector is constructed as $-\sum_i \vec p_T(i)$, where $i$ runs over hadrons, leptons and photons within the same $\eta$ ranges (the resulting missing energy is comparable to both truth- and jet-level missing energies constructed from the same objects).   Prompt leptons and those from tau decay are treated as lepton candidates, provided they survive isolation.   Jets that match a lepton within $\Delta R < 0.2$ and carry less than twice the lepton $p_T$ are discarded.  Following this, we apply a naive geometric isolation requirement to leptons, so that leptons within $\Delta R<0.4$ of higher-$p_T$ jets are rejected.  Finally, we apply a parametrized identification 
and reconstruction efficiency to lepton objects.  Electrons and muons are treated alike, with a parametrized efficiency that reflects their average.  Similarly, $b$-tags are applied by first matching each $b$-parton in an event with $|\eta|<3$ to its nearest jet, then applying a parametrized efficiency.   

For comparison, we also use PGS \cite{PGS} to establish a rough estimate of systematic errors 
involved in mocking-up a given set of object-level selections. 

The procedure above is clearly not  a faithful reproduction of the actual reconstruction performed by LHC experiments, nor is it intended to be.  It has been designed to model new-physics signals and physics backgrounds simply.  For example, no attempt is made to model resolution-dependent tails, fakes, or other instrumental effects. We have compared signal efficiencies obtained with this mockup with those reported in several new-physics searches at 36 pb$^{-1}$ at ATLAS and CMS, and in most cases they are compatible (after combining $e$ and $\mu$ channels) within about 10\%.  

\section*{Acknowledgments}
We thank Ilaria Segoni and Vitaliano Ciulli for very useful conversations about the CMS W+jets study.  
We thank Joshua Ruderman for comments and for essential help in checking the mock-ups of jets+$\MET$ analyses in ATLAS and CMS.  
We also thank Johan Alwall, Claudio Campagnari, Emmanuel Katz, Mariarosaria D'Alfonso, Joe Incandela, Eder Izaguirre, Sue Ann Koay, Michelangelo Mangano, Albert de Roeck, Roberto Rossin, Gavin Salam, Peter Skands, David Stuart, and Jay Wacker for valuable discussions.   M.L. acknowledges support from the Simons Postdoctoral Fellowship Program and the LHC Theory Initiative.  Research at the Perimeter Institute is supported in part by the Government of Canada through NSERC and by the Province of Ontario through MEDT.  The work of M.J.S. was supported by NSF grant PHY-0904069 and by DOE grant DE-FG02-96ER40959.  This research was supported in part by the National Science Foundation under Grant No. NSF PHY05-51164.     
	
\bibliography{multiplicitySearches}
\end{document}